\documentclass{SSA-BSSA}

\citethisauthor{T. Shimoda, W. Kokuyama, and H. Nozato}
\doi{00.0000/0000000000}
\recdate{00 Month 0000}

\begin{document}

%\title{Variation of sensitivity with temperature in broadband seismometers}
\title{Temperature dependence of broadband seismometer sensitivity}

\author[*1\orc{0000-0001-6198-1816}]{Tomofumi Shimoda}%
\author[1\orc{0000-0002-3843-2883}]{Wataru Kokuyama}
\author[1\orc{0000-0001-9614-2203}]{Hideaki Nozato}

\affil[1]{National Metrology Institute of Japan, National Institute of Advanced Industrial Science and Technology, Tsukuba, Japan}{\auorc{0000-0001-6198-1816}{(TS)}\auorc{0000-0002-3843-2883}{(WK)}\auorc{0000-0001-9614-2203}{(HN)}}
\corau{*Corresponding author: tomofumi.shimoda@aist.go.jp}

\begin{abstract}
Broadband seismometers are widely used in global observation networks deployed for geophysics research.
Recently, the calibration of their sensitivity has become an important factor for ensuring observation accuracy.
One of the limiting factors of calibration uncertainty is the temperature dependence of the sensitivity because seismometers operate in a wide range of temperatures. 
However, systems that accurately measure the temperature coefficient in seismometers have not been established.
Herein, we develop such a system using a triaxial vibration exciter combined with a thermostatic chamber.
Using this system, we calibrated several broadband seismometers (Trillium Compact, Trillium Horizon 360, and CMG-3T) at various temperatures, ranging from $-15~^\circ$C to $+45~^\circ$C.
We found that the temperature coefficient was $(0.11\pm0.01)$~\%/$^\circ$C, which can be attributed to the internal magnet of the seismometers.
The variation of the phase delay corresponded to a change of less than 2~ms in output time delay.
Additionally, we found that the relative frequency response below 1~Hz was stable against temperature variations.
These values are useful for evaluating the measurement accuracy of seismometer observation networks.
\end{abstract}

\maketitle

\begin{keypoints}
%\item
%Temperature coefficient of broadband seismometer sensitivity was measured.
\item
The sensitivities of three broadband seismometer models exhibited linear variation with temperature, with a coefficient of $(0.11\pm0.01)$~\%/$^\circ$C.

\item
Variations of the output delay over $-15~^\circ$C to $+45~^\circ$C were within 2~ms.

\item
Variations of the low-frequency cutoff were within 1~\%.

\end{keypoints}

\section{Introduction}
% 校正精度1%の要求
% 校正の先行研究：計量・地質それぞれ
% 温度依存性は「考慮すべき」とされているが、測れていない状況
% 温度依存性は校正の不確かさにとっても重要だが、その後観測点に設置したのちの精度にとっても重要
Evaluating the measurement accuracy of broadband seismometers is important for ensuring the quality of observation data and detecting instrument failure.
%For example, the required calibration accuracy of the Global Seismographic Network (GSN) is 1~\% (\cite{Lay2002}).
For example, the Global Seismographic Network (GSN) requires a calibration accuracy of 1~\% as part of its design goal (\cite{Lay2002}).
In particular, calibrating the sensitivity under the operational state is essential.
Such calibration techniques have been developed by both the geophysical and metrological research communities.
In some calibration techniques, global events, such as normal modes (\cite{Davis2005}) or tides (\cite{Davis2007}), have been used as reference signals, with globally common or predictable amplitudes.
Another approach compares the observed earthquake signal with the expected vibration calculated using a specific model (\cite{Kimura2015,Ekstrom2018}).
The most direct approach is to compare a local vibration signal with the reference seismometer located near the target seismometer (\cite{Anthony2017,Klaus2024,Kokuyama2025}).
In these studies, the temperature coefficient of the seismometer sensitivity has been considered an important uncertainty factor that affects both the calibration accuracy of the reference seismometer and the stability of the target seismometer during observations.
However, accurate methods for evaluating the temperature coefficient in seismometer sensitivity have not been established.

% 振動センサの温度依存性：DCに感度を持つ加速度センサであれば、地球重力で評価可能
% DC感度を持たない地震計では、振動を加えながら温度を変える必要がある
% 1軸ではそのような装置があるが、地震計は3軸（かつ取り付け方向を変えられない）
Although a method for measuring the temperature coefficient of accelerometers have been standardized in ISO~16063-34 (\cite{ISO16063-34}), the corresponding measurement system has been designed for single-axis measurements.
Therefore, this method cannot be employed in broadband seismometers, which has three sensing axes, and their installation orientation is fixed.
Additionally, the size and weight of typical broadband seismometers ($\sim30$~cm and $\sim10$~kg) are larger than those of typical accelerometers; therefore, the capacity of the systems designed to measure the temperature coefficient in accelerometers may be insufficient for broadband seismometers.
A recent study (\cite{George2024}) adopted a temperature testbed approach; the target seismometer was placed inside an insulated box, and the vibration amplitudes of regional earthquakes were compared at different temperatures.

% そこで最近開発した3軸加振器+恒温槽を使って、3軸の依存性を実測した
% 中間帯域感度の絶対校正を行った
% 周波数特性について、calibration coilを使って評価を行った
In this study, based on a recently developed environmental testing system (\cite{Shimoda2025}), we propose and develop a method for measuring the temperature coefficient in broadband seismometer sensitivity.
Our system combines a triaxial vibration exciter with a thermostatic chamber and applies vibrations to a seismometer at various temperatures. 
With this configuration, the characteristics of various seismometers can be evaluated systematically within a short period. 
In the measurement, the output signal of the seismometer is compared with that of reference accelerometers to calculate the absolute midband sensitivity at each temperature.
Furthermore, the relative frequency response of the seismometer is also measured using calibration coils.
These measurements revealed that the temperature coefficient is commonly about $(0.11\pm0.01)$~\%/$^\circ$C, across different types of broadband seismometers, and that the low‑frequency cutoff in the millihertz range is stable against temperature variation.
Additionally, we quantitatively characterized the temperature dependence of the phase shift or the calibration coil coefficients.

\section{Measurement method} \label{sec:method}
\begin{figure} % fig1
\centering
\includegraphics[width=\columnwidth]{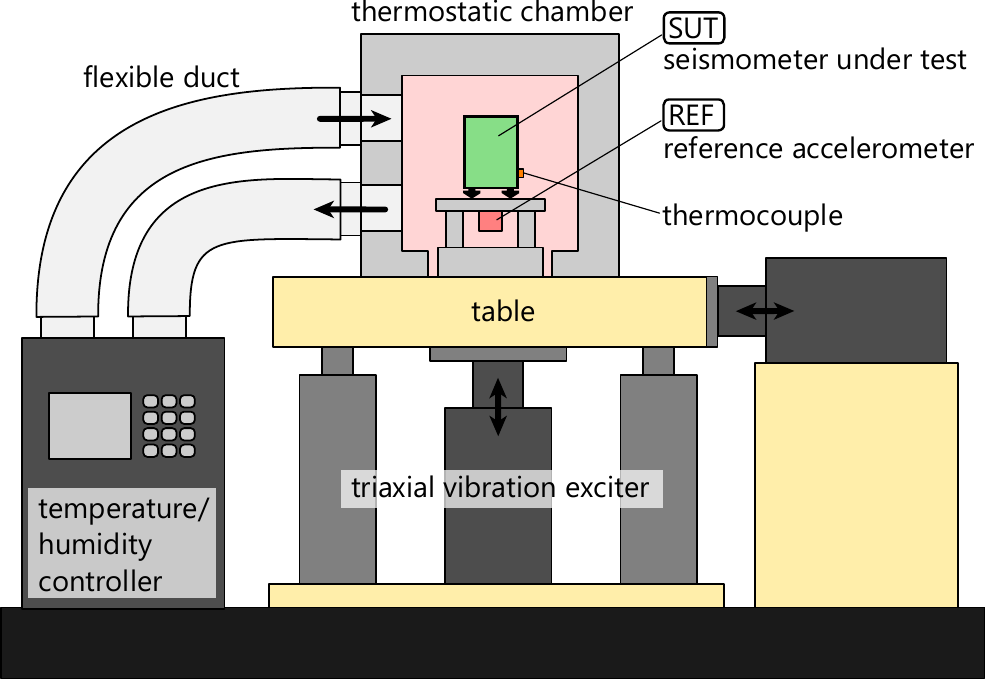}
\caption{Measurement setup for temperature coefficient of seismometer sensitivity.}
\label{fig:setup}
\end{figure}

Figure~\ref{fig:setup} shows the experimental setup.
A thermostatic chamber was fixed on a triaxial vibration exciter, and the seismometer under test (SUT), along with reference accelerometer (REF), was placed inside the chamber.
The system applied vibrations along the three orthogonal axis; the temperature inside the chamber varied from $-30~^\circ$C and $+80~^\circ$C.
In this study, the temperature points were set to $-15~^\circ$C, $0~^\circ$C, $+15~^\circ$C, $+23~^\circ$C, $+30~^\circ$C, and $+45~^\circ$C.
The setup was left overnight to allow the SUT temperature to stabilize at each temperature point; then, absolute midband sensitivity calibrations and relative frequency response measurements were conducted at each temperature.
The actual SUT temperature was monitored using a thermocouple (with a 1-$^\circ$C uncertainty) attached to the SUT.
The power supply, amplifier, and digitizers for the SUT and REF were placed outside the chamber.

Four broadband seismometers were tested as SUTs: two Trillium Compact (TC120, Nanometrics Inc.), a Trillium Horizon 360 (TH360, Nanometrics Inc.), and a CMG-3T (G\"{u}ralp Systems Ltd.). 
Note that the relative frequency response measurement was conducted for one of the TC120s because of the limited number of record channels in our digitizer.

\subsection{Absolute midband sensitivity calibration}

To calibrate the absolute midband sensitivity of the SUT, we applied sinusoidal vibrations with a frequency in the 1~Hz--5~Hz frequency range.
The output voltage signals of the SUT and REF were recorded using a PXIe-4464 (National Instruments) digitizer, which has an input impedance of 1~M$\mathrm{\Omega}$.
Then, we calculated the sensitivity of the SUT by comparing the output signal amplitudes as follows: 
\begin{equation}
S_\mathrm{SUT}(f) = \frac{ (2\pi i f) \tilde{V}_\mathrm{SUT}}{\tilde{V}_\mathrm{REF}} S_\mathrm{REF}(f),
\label{eq:Scal}
\end{equation}
where $\tilde{V}_\mathrm{SUT}$ and $S_\mathrm{SUT}$ are the complex amplitudes of the output signal (at frequency $f$) and the absolute sensitivity (in V/(m$\cdot$s$^{-1}$) unit) of the SUT, respectively.
$\tilde{V}_\mathrm{REF}$ and $S_\mathrm{REF}$ are the corresponding parameters of the REF, where $S_\mathrm{REF}$ is expressed in V/(m$\cdot$s$^{-2}$) units; thus, a conversion factor $(2\pi i f)$ was used to convert acceleration to velocity; this factor is included in the above equation.
Finally, the midband sensitivity $S_\mathrm{mid}$ is defined in this study as the average value of $|S_\mathrm{SUT}(f)|$ from 1~Hz to 1.6~Hz, where the frequency response is flat.

In this study, we used three servo accelerometers in $x$-, $y$-, and $z$-axes, respectively, as REFs.
Their absolute sensitivities $S_\mathrm{REF}$ and their temperature dependence were calibrated in advance by applying sinusoidal vibrations and comparing them with a reference laser interferometer included in the measurement system, following a primary calibration method standardized in ISO~16063-11 (\cite{ISO16063-11}).
In the calibration, the REF temperature was varied from $-20~^\circ$C and $+75~^\circ$C, and at each temperature the sensitivity was determined by the ratio of the REF output to the reference acceleration measured by the interferometer.
Since the interferometer was placed outside the chamber and only its laser beam was introduced into it, the calibration results reflect solely the temperature dependence of the REFs.  
The measured temperature coefficient of the REFs ($\sim0.008$~\%/$^\circ$C) was corrected during their comparison with the SUT.
Estimated relative standard uncertainty of the reference sensitivity $(u(|S_\mathrm{REF}|)/|S_\mathrm{REF}|)$ was 0.15~\%.
The main uncertainty factors are the accuracy of the voltage measurement, temperature fluctuation, the correction accuracy of the temperature dependence, and reproducibility.
Details of the calibration process and its uncertainty have been reported in \cite{Shimoda2025}.

To verify the calibration results of the SUT, we conducted validation experiments.
Initially, instead of the SUT, we evaluated the same type of servo accelerometer as that used as the REFs.
The measured temperature coefficient was 0.004~\%/$^\circ$C, which is similar orders of magnitude as the REFs (0.008~\%/$^\circ$C), demonstrating that the measurement procedure is appropriate.
Furthermore, in addition to the comparison calibration using the REFs (Eq.~(\ref{eq:Scal})), a primary calibration along $y$-axis was conducted for Trillium Horizon 360 using the laser interferometer, which works as a temperature-independent absolute reference as explained in the previous paragraph.
The results of the primary calibration were consistent with those of the comparison calibration within 0.1~\%, proving that the REFs did not introduce significant errors.
Based on these validations, we concluded that the results shown in the next section are the actual characteristics of the SUTs.

\subsection{Relative frequency response measurement}
The relative frequency response was measured using a calibration coil.
The coil and its paired magnet generate an electromagnetic force that simulates the inertial force caused by the ground acceleration. 
A sinusoidal voltage in the 4~mHz--1~Hz frequency range was applied to the calibration coil, and the equivalent acceleration was calculated using the coefficient of the coil $G_\mathrm{coil}$ (in (m$\cdot$s$^{-2}$)/V units), which was obtained from its datasheet.
Then, the sensitivity was calculated as follows:
\begin{equation}
S_\mathrm{SUT, coil}(f) = \frac{ (2\pi i f) \tilde{V}_\mathrm{SUT}}{G_\mathrm{coil} \tilde{V}_\mathrm{coil}} ,
\end{equation}
where $\tilde{V}_\mathrm{coil}$ is the complex amplitude of the calibration input voltage.
Again, factor $2\pi i f$ was used to convert acceleration to velocity.
The normalized value of $S_\mathrm{SUT, coil}(f)$ with respect to 1~Hz was calculated as the relative frequency response, which was then fitted to the following transfer function:
\begin{equation}
H(f) = \frac{ (2\pi i f)^2 }{ (2\pi i f + \alpha + i \beta)(2\pi i f + \alpha - i \beta) }.
\end{equation}
The frequency response of each SUT around the low-frequency pole is given in the datasheet by the above equation; $\alpha$ and $\beta$ are fitting parameters, which were used to calculate the cutoff frequency $f_\mathrm{c}=\sqrt{\alpha^2+\beta^2}/(2\pi)$.

Additionally, by comparing $S_\mathrm{SUT, coil}$ with the absolute sensitivity $S_\mathrm{mid}$, the absolute value of the coil coefficient $G_\mathrm{coil}$ can be derived as follows:
\begin{equation}
G_\mathrm{coil} = \frac{(2\pi i f) \tilde{V}_\mathrm{SUT}}{S_\mathrm{mid} \tilde{V}_\mathrm{coil}}. 	\label{eq:Gcoil}
\end{equation}
The comparison was performed using the data in the 1~Hz--1.6~Hz frequency range.
We evaluated the temperature coefficient of $G_\mathrm{coil}$.

\section{Measurement results}	\label{sec:result}
\begin{figure}
\centering
\begin{minipage}{0.9\columnwidth}
	\centering
	\includegraphics[width=\columnwidth]{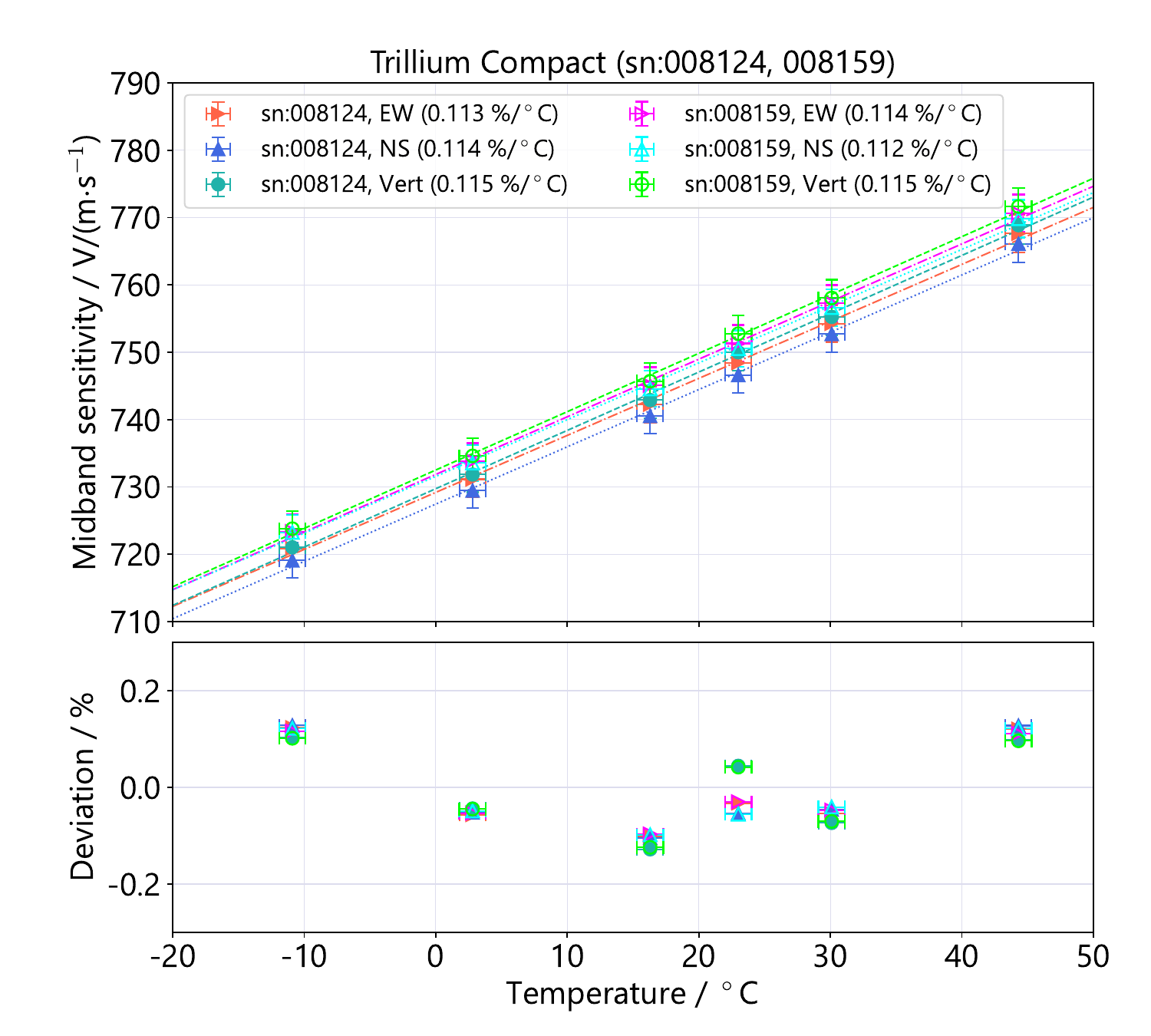}
\end{minipage}\\
\begin{minipage}{0.9\columnwidth}
	\centering
	\includegraphics[width=\columnwidth]{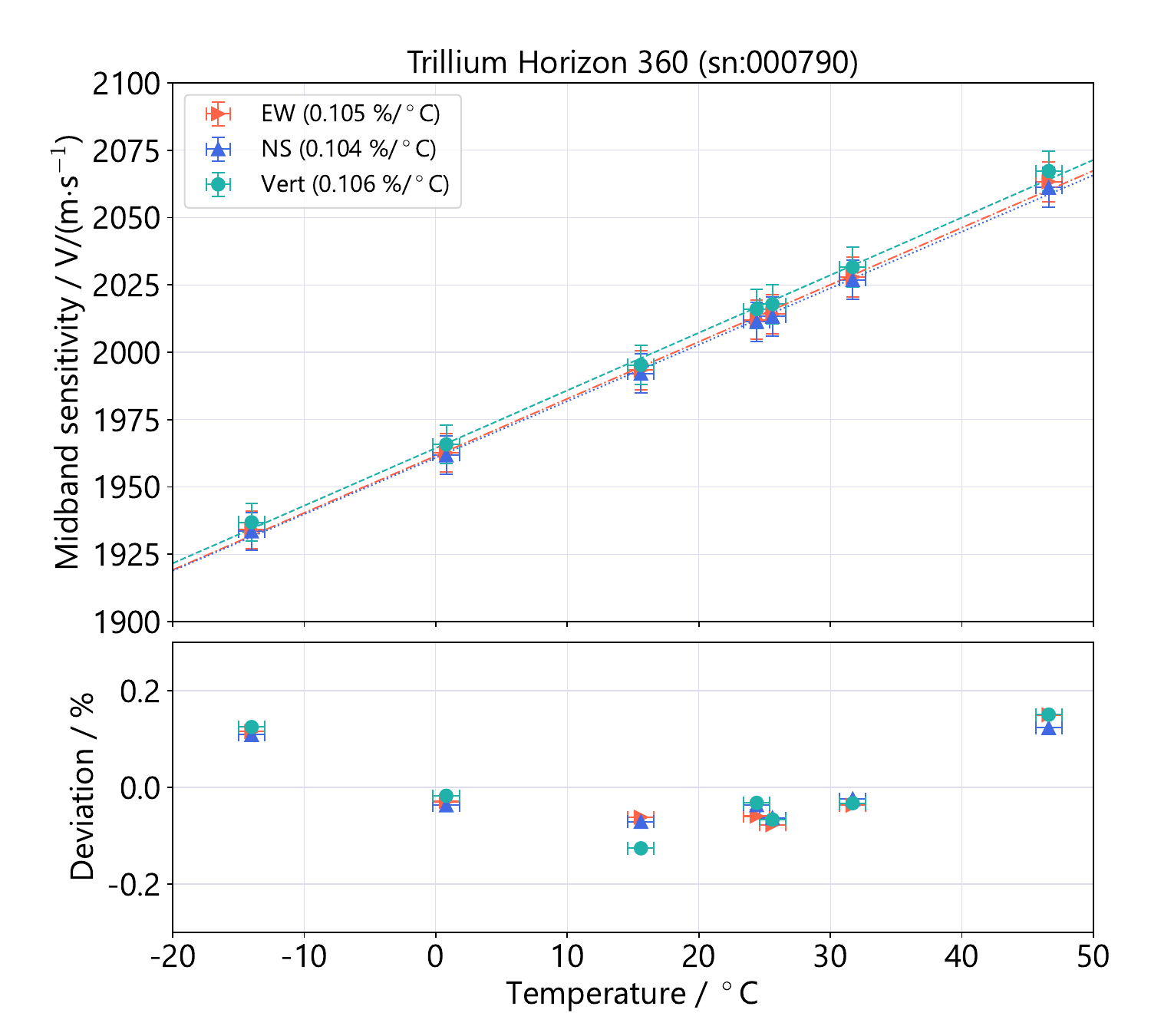}
\end{minipage}\\
\begin{minipage}{0.9\columnwidth}
	\centering
	\includegraphics[width=\columnwidth]{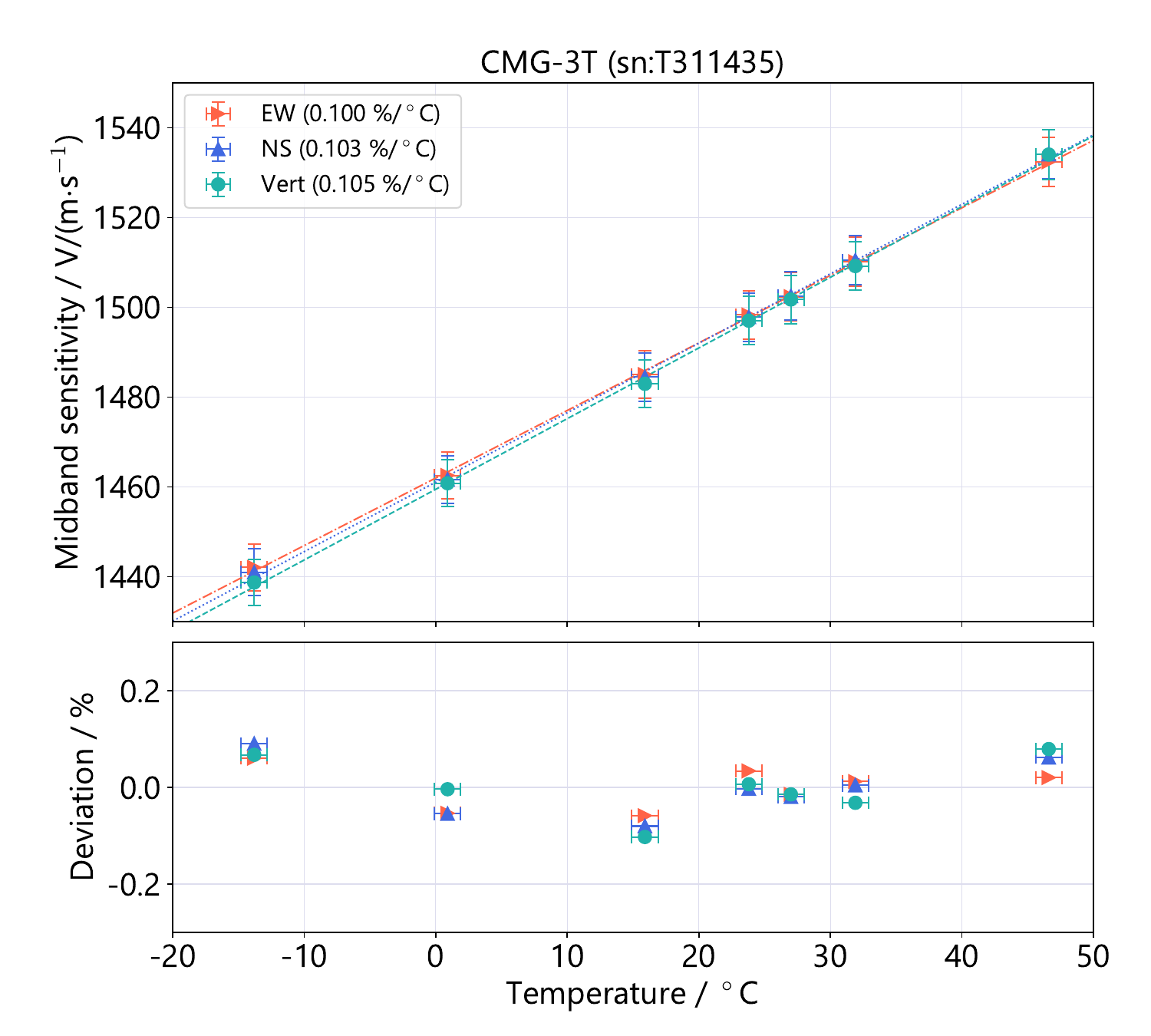}
\end{minipage}
\caption{Variation of the midband sensitivity with temperature of the Trillium Compact (top), Trillium Horizon 360 (center), and CMG-3T (bottom) seismometers. The upper half of each graph shows the measured values (points) and their linear fitting (lines). The lower half shows the deviation from the fitting.}
\label{fig:midband}
\end{figure}

\begin{table}
\tbl{Measured temperature coefficients\label{table:midband}}
{\begin{tabular*}{\columnwidth}{@{\extracolsep{\fill}}lllll@{}} %
	\multicolumn{2}{l}{\textbf{Sensor}}	&	\multicolumn{3}{c}{\textbf{Coefficients (\%/$^\circ$C)}}\\
	\cline{1-2}									\cline{3-5}	\\[-8pt]
	\multicolumn{1}{l}{\textbf{Model}}	&	\multicolumn{1}{l}{\textbf{Serial No.}}	&	\multicolumn{1}{l}{\textbf{EW}}	&	\multicolumn{1}{l}{\textbf{NS}}	&	\multicolumn{1}{l}{\textbf{Vert}}\\
	\colrule
	Trillium Compact 		& 008124		& 0.113	& 0.114	& 0.115		\\
	Trillium Compact		& 008159		& 0.114	& 0.112	& 0.115		\\
	Trillium Horizon 360	& 000790		& 0.105	& 0.104	& 0.106		\\
	CMG-3T 				& T311435	& 0.100	& 0.103	& 0.105		\\
	\botrule
\end{tabular*}}
{}
\end{table}

Figure~\ref{fig:midband} shows the measured midband sensitivities.
The estimated standard uncertainty of the sensitivity at each temperature was $\sim0.18$~\%, which is dominated by the relative uncertainty of the reference accelerometer sensitivity (0.15~\%), the contribution of temperature fluctuation during the measurement (0.1~\%), and the voltage measurement accuracy of the digitizer (0.02~\%).
As shown in Figure~\ref{fig:midband}, the variation of sensitivity with temperature is almost linear for each seismometer.
The fitted temperature coefficients are shown in the legend of Figure~\ref{fig:midband} and summarized in Table~\ref{table:midband}.
The coefficients of the four measured seismometers were similar ($\sim0.11$~\%/$^\circ$C); this value is much higher than that of the reference servo accelerometers ($\sim0.008$~\%/$^\circ$C).
The deviation from the linear fitting shows a small contribution of the higher-order temperature dependency of the sensitivity within $\pm0.15$~\%.

\begin{figure}
\centering
\begin{minipage}{0.9\columnwidth}
	\centering
	\includegraphics[width=\columnwidth]{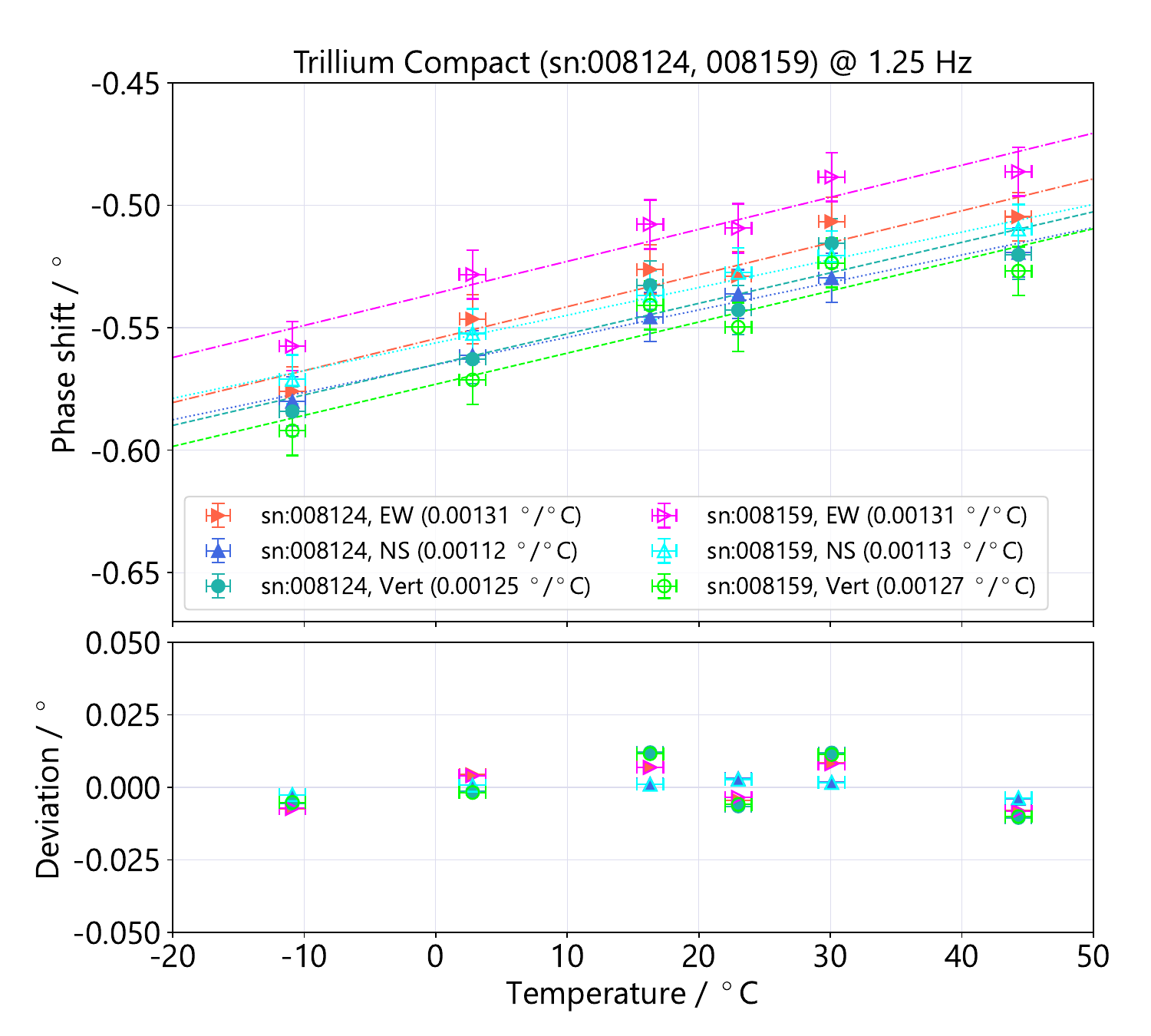}
\end{minipage}\\
\begin{minipage}{0.9\columnwidth}
	\centering
	\includegraphics[width=\columnwidth]{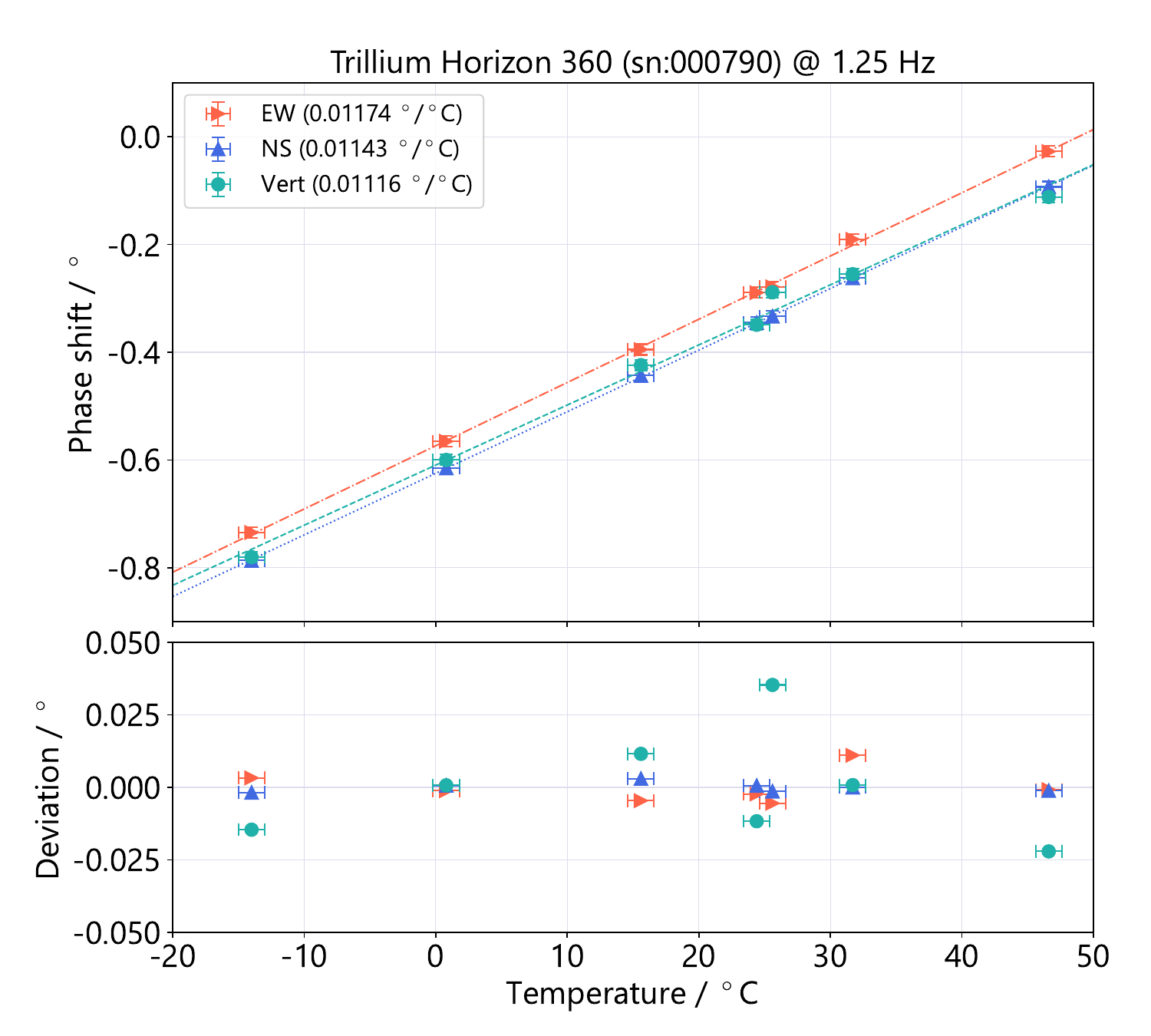}
\end{minipage}\\
\begin{minipage}{0.9\columnwidth}
	\centering
	\includegraphics[width=\columnwidth]{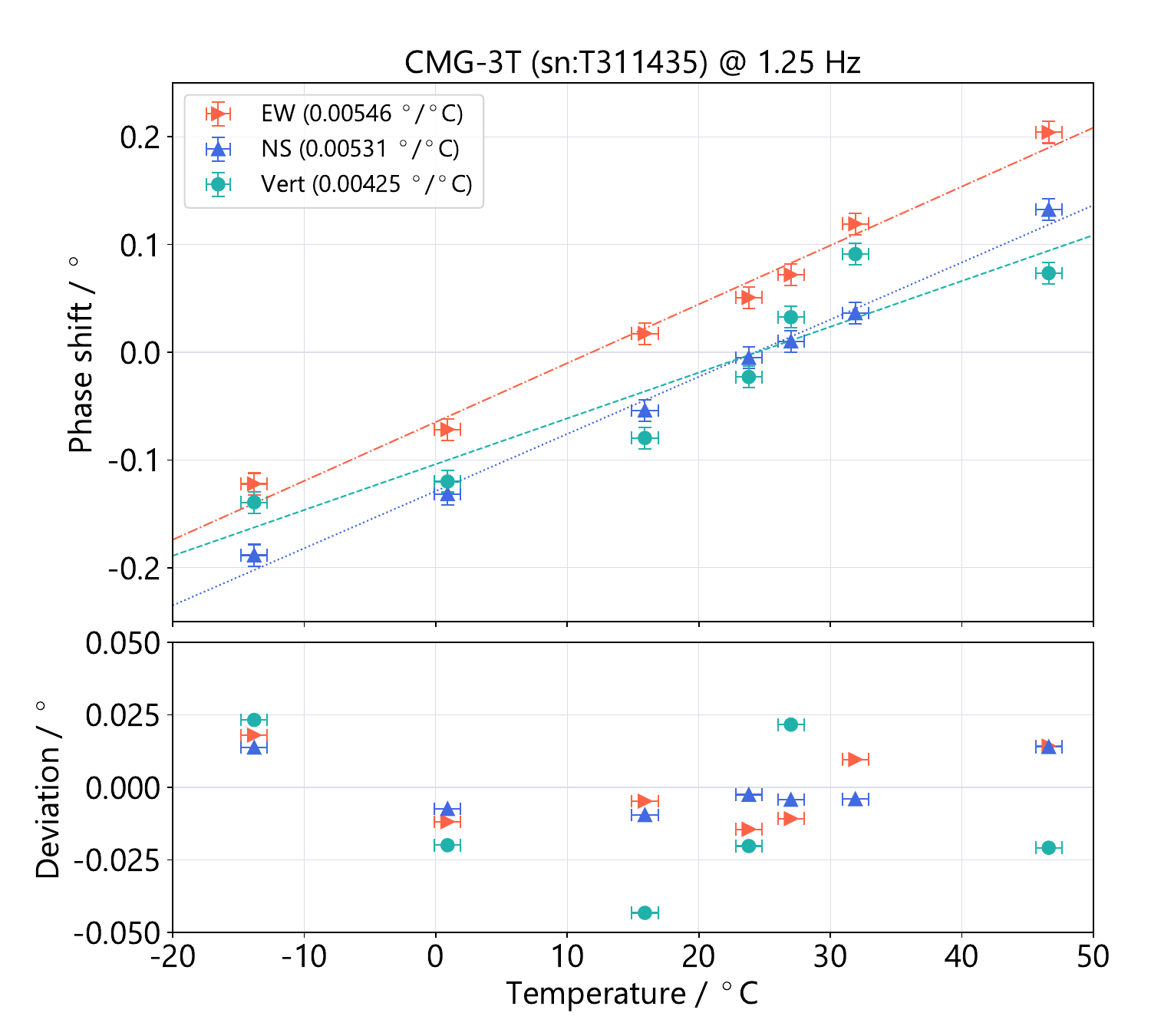}
\end{minipage}
\caption{Variation of the phase shift with temperature at 1.25~Hz of the Trillium Compact (top), Trillium Horizon 360 (center), and CMG-3T (bottom) seismometers. }
\label{fig:phase}
\end{figure}

\begin{figure}
\centering
\begin{minipage}{1\columnwidth}
	\centering
	\includegraphics[width=\columnwidth]{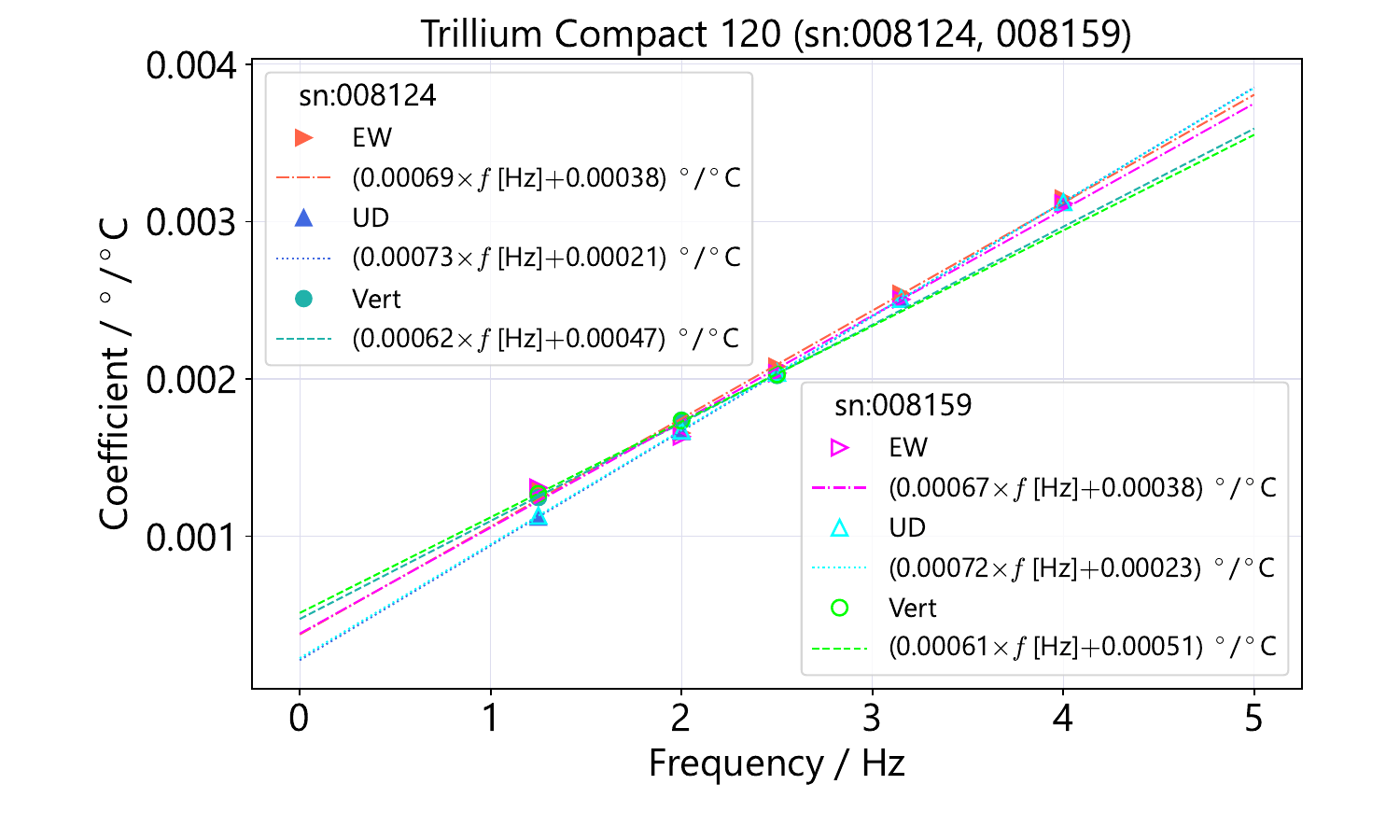}
\end{minipage}\\
\begin{minipage}{1\columnwidth}
	\centering
	\includegraphics[width=\columnwidth]{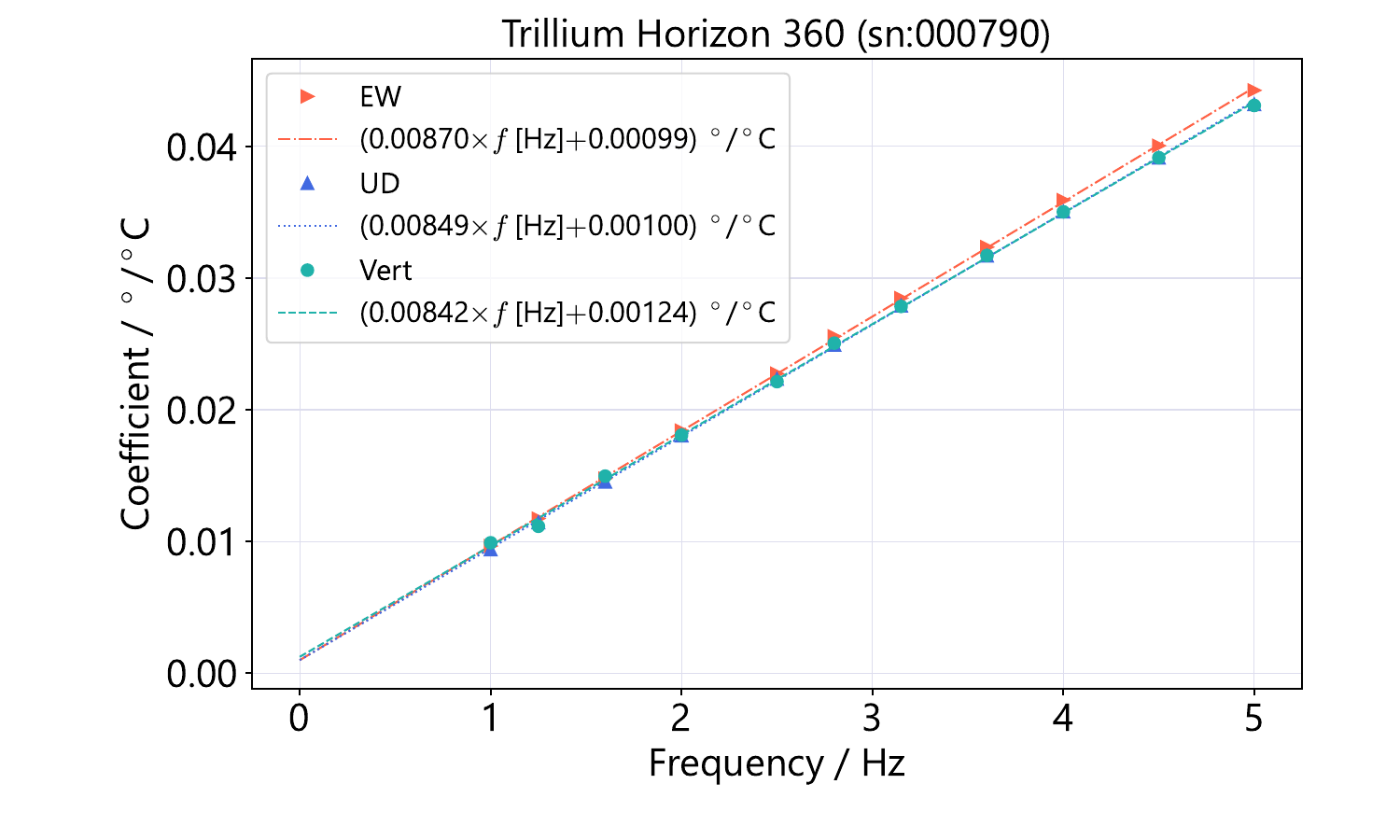}
\end{minipage}\\
\begin{minipage}{1\columnwidth}
	\centering
	\includegraphics[width=\columnwidth]{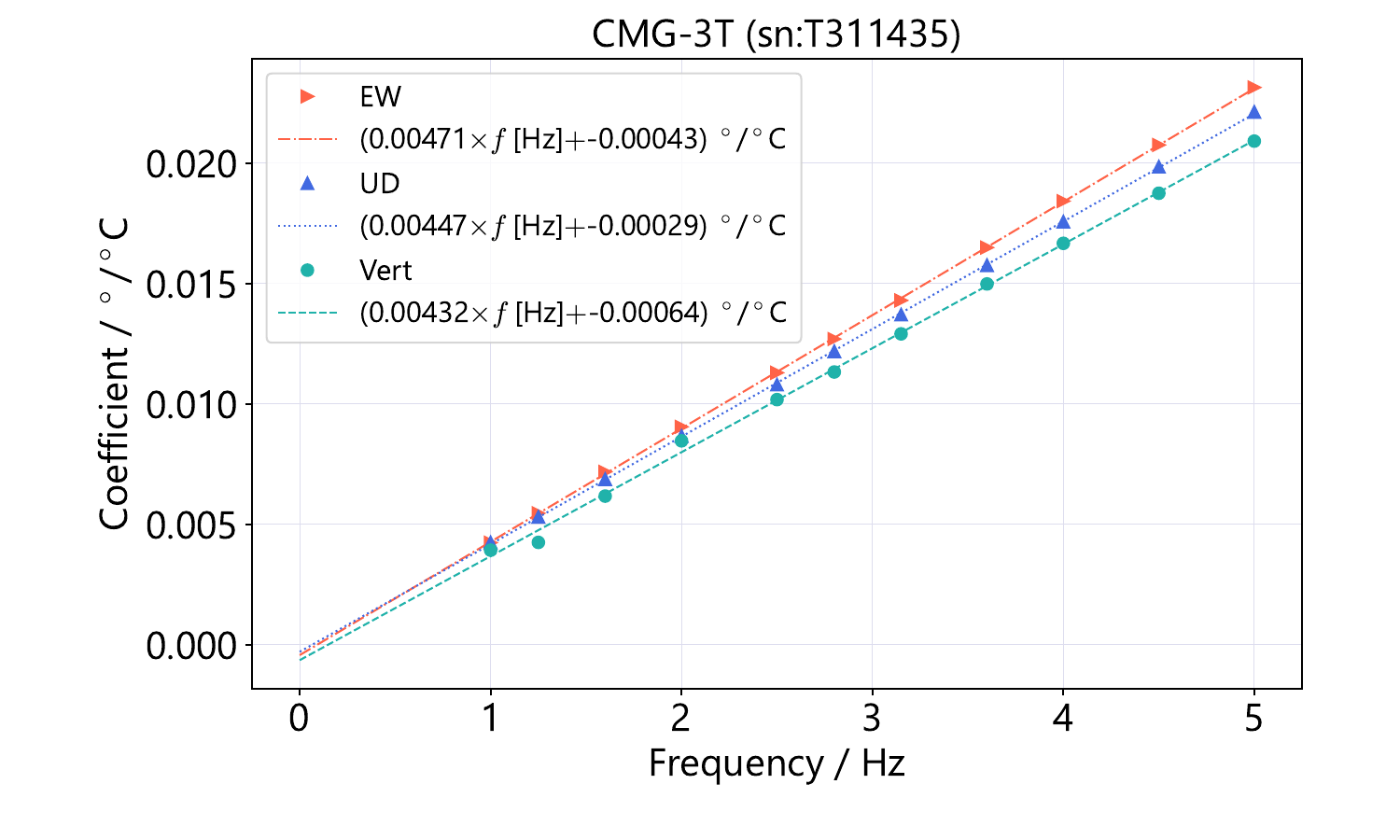}
\end{minipage}
\caption{Frequency dependence of the temperature coefficient of phase shift of the Trillium Compact (top), Trillium Horizon 360 (center), and CMG-3T (bottom) seismometers. }
\label{fig:phase-coefficient}
\end{figure}
Figure~\ref{fig:phase} shows the temperature dependence of the phase shift, $\arg(S_\mathrm{SUT})$.
Since the phase shift depends on frequency, the result at 1.25~Hz is presented as a representative value.
As with the midband sensitivity, the phase shift exhibited a linear dependence on temperature with coefficients of approximately 0.001$^\circ/^\circ$C, 0.011$^\circ/^\circ$C, and 0.005$^\circ/^\circ$C at 1.25~Hz for the Trillium Compact, Trillium Horizon 360s, and CMG-3T, respectively. 
The frequency dependence of the temperature coefficients is plotted in Figure~\ref{fig:phase-coefficient}.
The coefficients are roughly proportional to frequency and become small in the low-frequency range.

%\begin{figure}
%\centering
%\begin{minipage}{1\columnwidth}
%	\centering
%	\includegraphics[width=\columnwidth]{TC120(sn008124)_electric_LF.pdf}
%\end{minipage}\\
%\begin{minipage}{1\columnwidth}
%	\centering
%	\includegraphics[width=\columnwidth]{TH360_electric_LF.pdf}
%\end{minipage}\\
%\begin{minipage}{1\columnwidth}
%	\centering
%	\includegraphics[width=\columnwidth]{CMG-3T_electric_LF.pdf}
%\end{minipage}
%\caption{Relative frequency response of Trillium Compact (top), Trillium Horizon 360 (center), and CMG-3T (bottom) at low frequencies. }
%\label{fig:LF}
%\end{figure}

\begin{figure}
\centering
\begin{minipage}{1\columnwidth}
	\centering
	\includegraphics[width=\columnwidth]{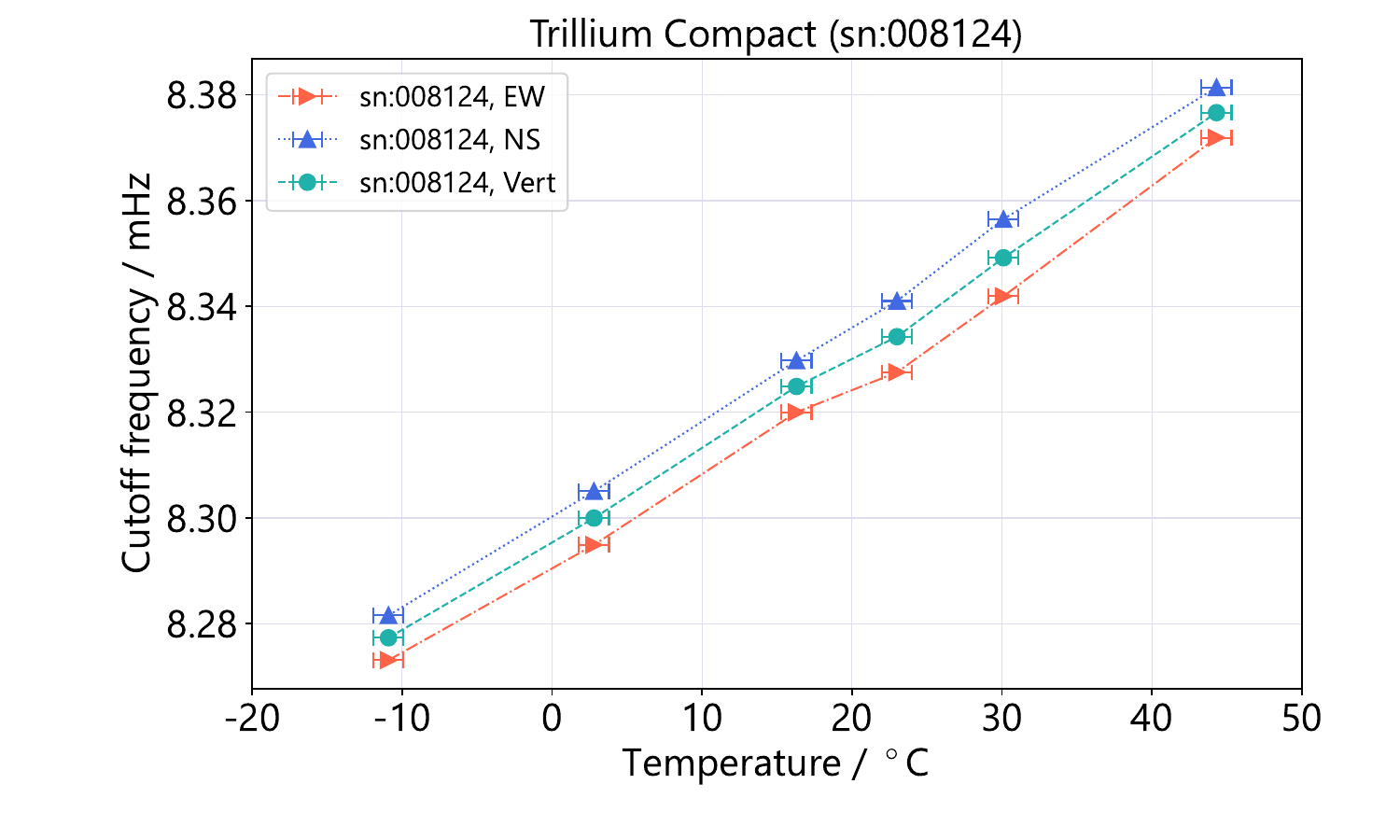}
\end{minipage}\\
\begin{minipage}{1\columnwidth}
	\centering
	\includegraphics[width=\columnwidth]{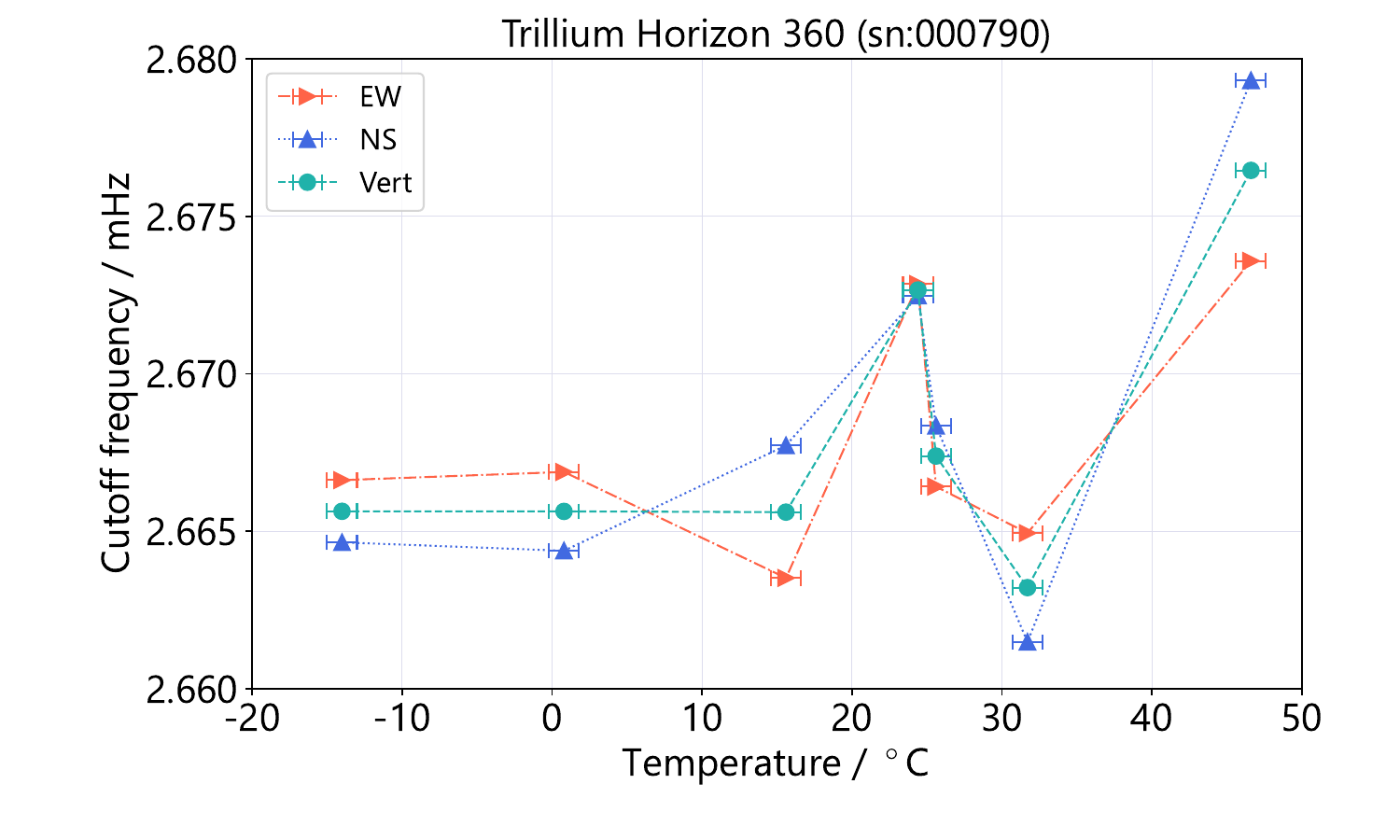}
\end{minipage}\\
\begin{minipage}{1\columnwidth}
	\centering
	\includegraphics[width=\columnwidth]{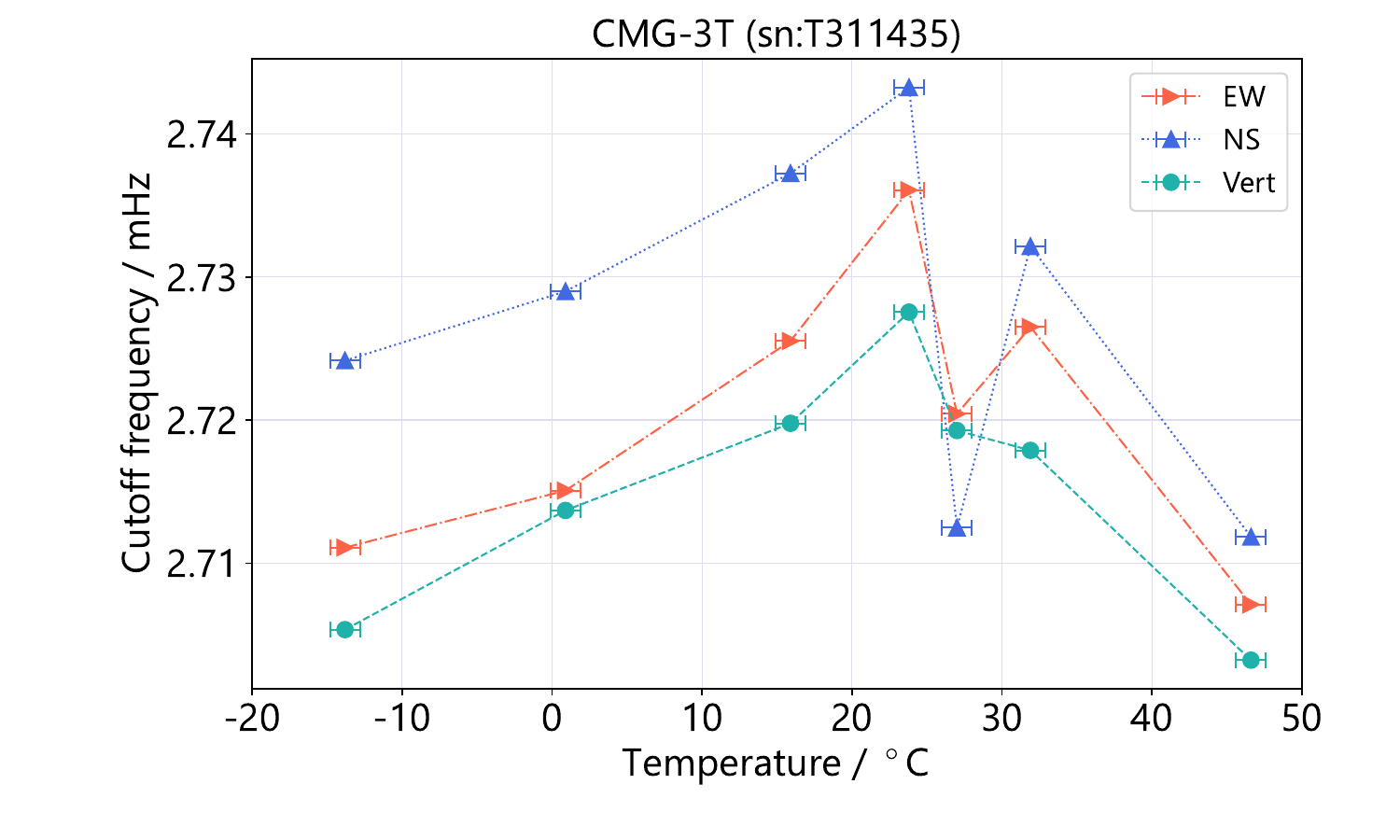}
\end{minipage}
\caption{Variation of the fitted low-frequency cutoff frequency with temperature of the Trillium Compact (top), Trillium Horizon 360 (center), and CMG-3T (bottom) seismometers. }
\label{fig:pole}
\end{figure}

Figure~\ref{fig:pole} shows the variation of the cutoff frequency fitted from relative frequency response at different temperatures.
%No significant change in the response was observed for any seismometers.
%For quantitative discussion, the temperature dependence of the fitted cutoff frequency is shown in Figure~\ref{fig:pole}.
The Trillium Compact seismometers exhibit a linear dependence with a coefficient of 0.002~mHz/$^\circ$C.
In contrast, no clear trend in the cutoff frequency variation is observed for the other two seismometers.
This might be explained by the large fitting errors caused by the cutoff frequency ($\approx2.7$~mHz) being outside the measurement frequency range ($>4$~mHz).
The deviation over the temperature range ($-15~^\circ$C to $+45~^\circ$C) was within approximately 1~\% of each cutoff frequency.

\begin{figure}
\centering
\begin{minipage}{1\columnwidth}
	\centering
	\includegraphics[width=\columnwidth]{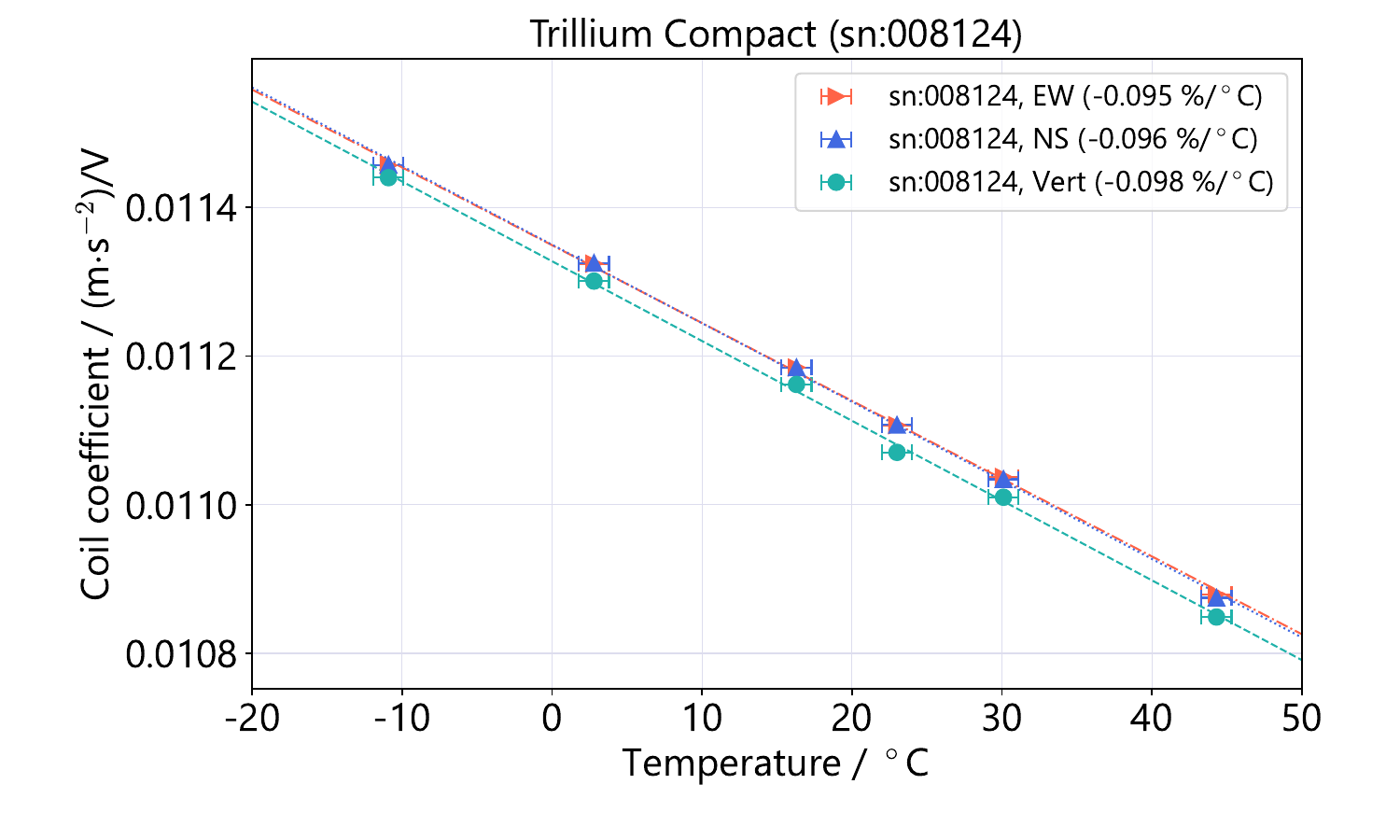}
\end{minipage}\\
\begin{minipage}{1\columnwidth}
	\centering
	\includegraphics[width=\columnwidth]{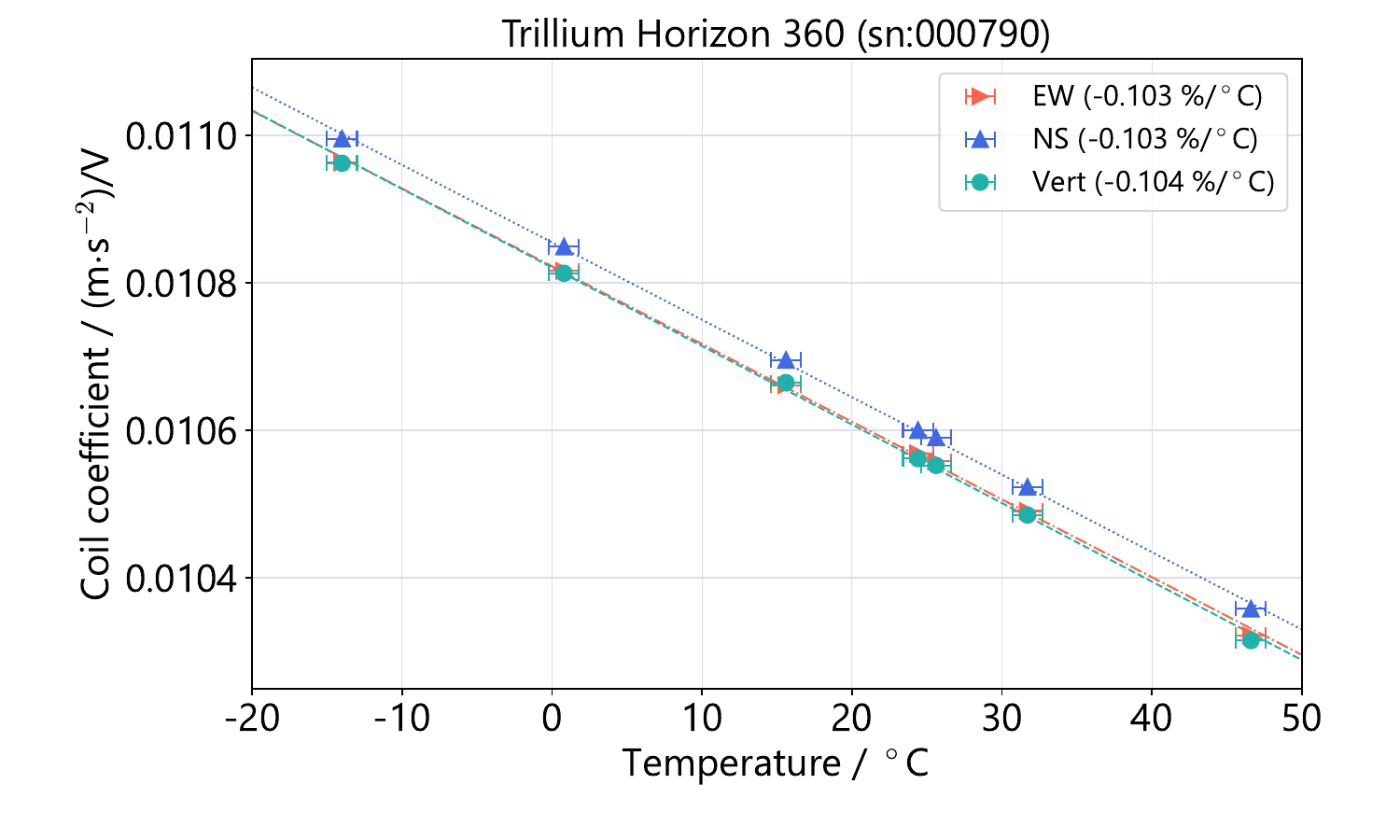}
\end{minipage}\\
\begin{minipage}{1\columnwidth}
	\centering
	\includegraphics[width=\columnwidth]{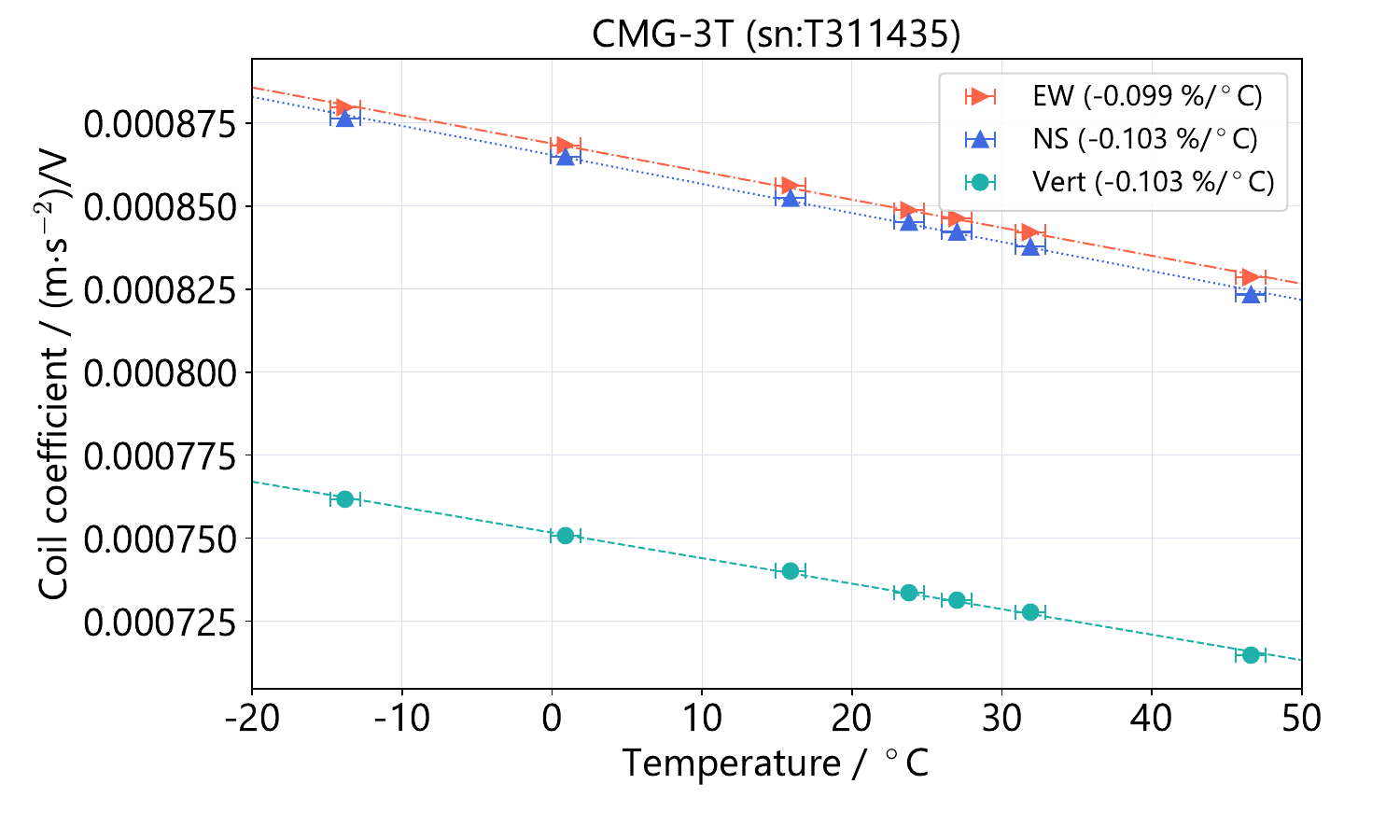}
\end{minipage}
\caption{Variation of the calibration coil coefficient $G_\mathrm{coil}$ with temperature of the Trillium Compact (top), Trillium Horizon 360 (center), and CMG-3T (bottom) seismometers. }
\label{fig:Gcoil}
\end{figure}

Figure~\ref{fig:Gcoil} shows the variation of the calibration coil coefficient with temperature calculated using Eq.~(\ref{eq:Gcoil}).
We observed that $G_\mathrm{coil}$ has inverse dependence compared with that of the midband sensitivity.
Consequently, the product $S_\mathrm{mid}\cdot G_\mathrm{coil}$ yields small temperature coefficients of 0.018~\%/$^\circ$C for the Trillium Compact seismometers and 0.002~\%/$^\circ$C for the Trillium Horizon 360 and CMG-3T seismometers.
These values indicate that variation in $G_\mathrm{coil}$ is the main source of the temperature coefficient in $S_\mathrm{mid}$, as discussed in the next section.

\section{Discussion}\label{sec:discussion}
% 感度の0.11%/degの依存性は無視できない
% 温度が10℃異なれば感度は1%変わり、
% globalに数％レベルの系統的な偏差（南極では振動が小さく見えるなど）が生じる可能性もある
The estimated temperature coefficient of the midband sensitivity was $(0.11\pm0.01)$~\%/$^\circ$C.
This value may not be negligible in seismic observations.
For example, a temperature difference of 10~$^\circ$C corresponds to a 1.1~\% difference in sensitivity, exceeding the accuracy required by the GSN (\cite{Lay2002}).
When a seismometer that has been calibrated in the laboratory is moved to an observation site, its sensitivity should be corrected according to the on-site temperature.
Additionally, the temperature coefficient can cause a global systematic difference in the observed vibration amplitude up to a few percent; for example, the temperatures around the equator and the Antarctic differ by a few tens of degrees.
Therefore, the temperature at each observation site must be considered to ensure observation accuracy.

\begin{figure} % fig1
\centering
\includegraphics[width=0.8\columnwidth]{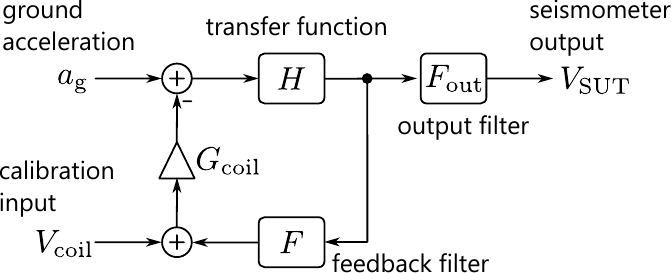}
\caption{Simplified block diagram of a typical force-balance seismometer. The transfer function $H$ includes a pendulum response, a displacement transducer, and an integrator.}
\label{fig:block_diagram}
\end{figure}
Next, we discuss the origin of the temperature dependence.
Figure~\ref{fig:block_diagram} shows a simplified block diagram of a typical broadband seismometer (based on \cite{Ackerley2015,CMGmanual}).
The velocity sensitivity is derived as follows:
\begin{equation}
S_\mathrm{SUT} = \frac{ (2\pi i f) V_\mathrm{SUT}}{a_\mathrm{g}} =  \frac{ (2\pi i f)F_\mathrm{out}H}{1+HFG_\mathrm{coil}} \approx \frac{ (2\pi i f)F_\mathrm{out}}{FG_\mathrm{coil}}.
\end{equation}
In the above equation, the sufficiently high ($HFG_\mathrm{coil}\gg 1$) was used for the approximation. 
As mentioned in the previous section, the product $S_\mathrm{SUT}\cdot G_\mathrm{coil}=(2\pi if)F_\mathrm{out}/F$ yields a small temperature coefficient. 
Therefore, the temperature coefficient of $S_\mathrm{SUT}$ or $S_\mathrm{mid}$ mainly originates from $G_\mathrm{coil}$.
The most probable cause of the variation in $G_\mathrm{coil}$ is the magnet that is paired with the coil; for example, the temperature coefficient of the residual magnetic flux density of an NdFeB magnet ranges from $-0.09$~\%/$^\circ$C to $-0.12$~\%/$^\circ$C (\cite{Stan2017}).
A similar conclusion was reported by \cite{Geoffrey2025}. 
However, further investigation requires detailed information on the component, which is beyond the scope of this study.

%位相シフトの温度依存性は地震計によってさまざまであった。
%－15℃から45℃にかけての、1.25Hzでの変動の幅はTC120が0.1°（0.2ms相当）、TH360が0.8°（1.7ms相当）、CMG-3Tが0.4°（0.9ms相当）であった（図５）。
%いずれも通常の広帯域観測では大きな問題にはならないレベルだろうと思われる。例えば、時刻精度に関してはGSN Design goalsで10msが目標にされている。
%いずれも共通して、温度係数が周波数に比例し、低周波に行くほど温度変化は小さかった（図４）。
%これは高周波側の特性変化の寄与が主要であることを示している。ただし詳しい由来は特定されていない。
The temperature coefficients of the phase shift varied among the seismometers.
At 1.25~Hz, the variations over the measured temperature range ($-15~^\circ$C to $+45~^\circ$C) were 0.1$^\circ$, 0.8$^\circ$, and 0.4$^\circ$ for the Trillium Compact, Trillium Horizon 360, and CMG-3T seismometers, respectively (Figure~\ref{fig:phase}).
The corresponding time shifts were 0.2~ms, 1.7~ms, and 0.9~ms.
Given the timing accuracy required by the GSN (10~ms), these variations will not seriously matter in broadband seismic observation.
For all seismometers, the temperature coefficient was approximately proportional to frequency (Figure~\ref{fig:phase-coefficient}).
This suggests that the variations of the phase shift with temperature may be attributable to high-frequency characteristics of mechanical components or electronics.

%mHz帯のカットオフ周波数は温度によってほとんど変化しなかった。
%この特性は内部のフィードバック回路によって決まっており、通常は抵抗値やキャパシタによって構成されている。
%温度係数は種類によって様々だが、典型的には100ppm/℃のオーダーである。
%従って今回の温度範囲（－15℃から45℃）では0.6％程度の変化と想定され、8.3mHzに対して0.05mHz、2.7mHzに対して0.02mHzのオーダーの変化が想定される。
%コンポーネントの温度係数はモノによって数倍の差があることを考慮すると、図5の結果は少なくともこれと矛盾しないものである。
The variation in the low-frequency cutoff was small.
In the broadband seismometers, the low-frequency cutoff is determined by the feedback circuits composed of resistors and capacitors.
Typically, these components have temperature coefficients on the order of 100~ppm/$^\circ$C, although the value may differ by several factors depending on the specific products.
Consequently, they can cause a frequency shift of approximately 0.6~\% over the temperature range ($-15~^\circ$C to $+45~^\circ$C), corresponding to 0.05~mHz for Trillium Compact or 0.02~mHz for Trillium Horizon 360 and CMG-3T seismometers.
Considering the uncertainty of the components' temperature coefficients, the results in Figure~\ref{fig:pole} are consistent with this estimation.

\section{Conclusion}\label{sec:conclusion}
Four broadband seismometers were calibrated at different temperatures.
The measured temperature coefficient of the midband sensitivity was $(0.11\pm0.01)$~\%/$^\circ$C, and the relative frequency response was stable against temperature variations.
This dependence should be considered when the seismometer operates at temperatures quite different from room temperature ($\sim 23$~$^\circ$C).
To ensure observation accuracy within 1~\%, the temperature should be monitored within 8-$^\circ$C accuracy.
This can be easily achieved using a proper thermometer.

The temperature coefficient of the permanent magnet used as a coil-magnet actuator probably causes a sensitivity variation with temperature; this variation is expected to be consistent across the same-model seismometers.
In fact, two Trillium Compact seismometers used in this study exhibited quite a similar temperature coefficients (Figure~\ref{fig:midband}).
This can simplify the correction of the temperature-dependent sensitivity because only small number of individuals for each model will need to be characterized.
As a next step, we will investigate more seismometers to establish the correction method.
Such work can improve the accuracy and reliability of global seismic observation networks.

%%%%%%%%%%%%%%%%%%%
%%%%%%%%%%%%%%%%%%%
%%%%%%%%%%%%%%%%%%%

\begin{datres}%[CUSTOM HEAD]
All measurement data used in this paper were collected by ourselves and are available upon reasonable request.
\end{datres}

\section{Declaration of Competing Interests}
The authors acknowledge that there are no conflicts of interest recorded.

\begin{ack}%[CUSTOM ACK HEAD]
This work was supported by a project commissioned by the New Energy and Industrial Technology Development Organization (NEDO), Japan. 
\end{ack}

\bibliography{seismometer.bib}

\end{document}